\documentclass[twocolumn]{aastex631}

\usepackage{newtxtext,newtxmath}

\usepackage{orcidlink}
\usepackage{float}
\graphicspath{{figures/}}

\defcitealias{Zou_2026}{Paper~I}
\defcitealias{Zou_2026II}{Paper~II}

\begin{document}

\title{A systematic study of AGN feedback in a disk galaxy using MACER. III. High Gas Fractions in AGN Hosts}

\author{Yuxuan Zou\orcidlink{0009-0006-4662-3053}}
\affiliation{Astrophysics Division, Shanghai Astronomical Observatory, Chinese Academy of Sciences, 80 Nandan Road, Shanghai 200030, China}
\affiliation{University of Chinese Academy of Sciences, No. 19A Yuquan Road, Beijing 100049, China}

\author{Feng Yuan\orcidlink{0000-0003-3564-6437}}
\affiliation{Center for Astronomy and Astrophysics and Department of Physics, Fudan University, Shanghai 200438, China}

\author{Suoqing Ji\orcidlink{0000-0001-9658-0588}}
\affiliation{Center for Astronomy and Astrophysics and Department of Physics, Fudan University, Shanghai 200438, China}
\affiliation{Key Laboratory of Nuclear Physics and Ion-Beam Application (MOE), Fudan University, Shanghai 200433, China}
\correspondingauthor{Feng Yuan, Suoqing Ji}
\email{fyuan@fudan.edu.cn,sqji@fudan.edu.cn}

\author{Jinyi Shangguan\orcidlink{0000-0002-4569-9009}}
\affiliation{Kavli Institute for Astronomy and Astrophysics, Peking University, Beijing 100871, China}

\author{Hassen M. Yesuf\orcidlink{0000-0002-4176-9145}}
\affiliation{Shanghai Astronomical Observatory, Chinese Academy of Sciences, 80 Nandan Road, Shanghai 200030, China}

\author{Lu Shen\orcidlink{0000-0001-9495-7759}}
\affiliation{Center for Astronomy and Astrophysics and Department of Physics, Fudan University, Shanghai 200438, China}

\author{Luis C. Ho\orcidlink{0000-0001-6947-5846}}
\affiliation{Kavli Institute for Astronomy and Astrophysics, Peking University, Beijing 100871, China}
\affiliation{Department of Astronomy, School of Physics, Peking University, Beijing 100871, China}

\begin{abstract}

We use high-resolution hydrodynamic simulations in the MACER framework to explain why low-redshift PG quasar hosts can retain substantial cold-gas reservoirs, with gas fractions and gas-to-stellar mass ratios showing little dependence on instantaneous AGN luminosity. This paper is the third in a series systematically studying AGN feedback in a disk galaxy subject to cosmological gas inflow. The simulations include multiphase gas, star formation, stellar feedback, and self-consistent radiative and mechanical AGN feedback. We reproduce the observed weak connection between host-galaxy gas content and AGN luminosity over  $L_{\rm AGN}/L_{\rm Edd}\sim 10^{-5}-10$, while the galaxy nevertheless undergoes pronounced gas depletion and star-formation quenching. During the quenching phase, the cold-gas mass declines by nearly three orders of magnitude, and the evolution of the gas distribution shows that AGN feedback progressively removes both cold and hot gas from the galaxy. The star formation rate is more closely linked to the cold-gas mass than to AGN luminosity. This behavior arises from a timescale mismatch: AGN luminosity varies on $\sim 10^5-10^6$ yr timescales, whereas repeated AGN-driven outflows cumulatively deplete the galaxy-scale gas reservoir over $\sim$ 1 Gyr. Our simulations therefore provide a physical explanation for the gas-rich PG quasar hosts and show that their observed gas properties are fully consistent with effective, long-term ejective AGN feedback.

\end{abstract}


\section{Introduction}

A central goal of galaxy evolution studies is to understand how supermassive black holes (SMBHs) and their host galaxies co-evolve. AGN feedback is often discussed in terms of three related but distinct concepts. \textit{Positive feedback} refers to processes in which nuclear activity enhances, rather than suppresses, local star formation \citep{Wagner2016}. Examples include AGN-driven compression or shocks that trigger or enhance star formation in specific regions, as supported by both observational and theoretical studies \citep{Maiolino_2017,Cresci_2018,Gallagher_2019,2026ApJ..1004..128S}.
The positive correlation between the star formation rate and black hole accretion rate is also sometimes referred to as positive feedback. However, this correlation may simply arise because a gas-rich, fueling-dominated phase can sustain both accretion onto the SMBH and star formation from a common gas reservoir. Because this scenario does not require the AGN to actively promote either process, it should not, strictly speaking, be termed ``feedback.''
{Consistent with this fueling interpretation, \citet{Shangguan_2020_ApJ_feedback} found that the \(L_{\rm AGN}\)--SFR correlation disappears after removing the common dependence on gas mass, whereas \citet{Zhuang_2021} reported a residual correlation that they interpreted as possible positive feedback. The reason for the discrepancy remains unclear, but it may be related to differences between their samples.}
\textit{Negative feedback}, by contrast, occurs when AGN radiation, winds, and jets deposit energy and momentum into the interstellar medium (ISM), thereby heating or expelling gas and reducing the fuel available for subsequent star formation and black hole growth. Negative feedback is widely invoked to regulate black hole growth, establish scaling relations such as the \(M\)--\(\sigma\) relation \citep{Kormendy_ho_2013,McConnell_2013}, and maintain massive galaxies in a quiescent state \citep{Fabian_2012,King2015,Harrison2018,2023MNRAS.525.4840Z}. Finally, observers sometimes report a \textit{null} (``zero-feedback'') result: the (cold) gas fractions of galaxies with different levels of AGN activity are similar, seemingly indicating that AGN feedback plays no significant role. In this paper, we investigate how this observation should be interpreted and, in particular, whether it conflicts with the negative-feedback scenario.

% Finally, observers sometimes draw a \textit{zero-feedback} conclusion from a single snapshot: if a galaxy still hosts abundant cold gas while the AGN is luminous, negative feedback is taken to be weak or absent. The main question addressed in this paper is whether that zero-feedback inference is justified, or whether it instead reflects a mismatch between rapidly changing nuclear luminosity and a galaxy-scale gas reservoir that evolves much more slowly.

In the negative-feedback scenario, AGN feedback can, at least in principle, remove a substantial fraction of the ISM from a galaxy and thereby suppress star formation \citep{silk1998quasars,King_2003,Murray_2005,Hopkins_2016,weinberger2018supermassive}. In the simplest picture, a quasar host undergoing effective negative feedback would therefore be expected to appear gas-poor while the AGN is active. Observations, however, paint a more complicated picture: although multiphase outflows have been detected in many AGN \citep{Cicone_2014,Zakamska_2014,Perna_2015,morganti_2016,Fiore_2017,Baron_2018,Fluetsch_2019}, substantial reservoirs of atomic and molecular gas remain common in AGN host galaxies \citep{Fabello_2011,Gereb_2015,Zhu_2015,Ellison_2018,Zhang_2019_ApJL,Zhang_2021_ApJ,Guo_2021,Guo_2022_ApJL}. {Consistently, H\,I surveys of nearby type~1 AGNs show that their hosts are generally at least as gas-rich as inactive galaxies of similar morphological type, and that the global atomic-gas content does not track the instantaneous nuclear luminosity or Eddington ratio \citep{Ho_2008}.
Statistical analyses of type~2 AGN hosts similarly show that strongly accreting systems remain gas-rich and star-forming, with no evidence that AGN feedback instantaneously depletes the molecular reservoir on \(\lesssim0.5\,\mathrm{Gyr}\) timescales \citep{Yesuf_2020}.}

For example, at low redshift, Palomar--Green (PG) quasar hosts are found to contain galaxy-scale gas reservoirs, as inferred from dust spectral energy distributions \citep{Shangguan_2018}, as well as molecular gas traced by CO \citep{Shangguan_2020,Shangguan_2020_ApJ_feedback,Shangguan_2019}. These gas reservoirs are broadly similar to those in massive star-forming galaxies, while PG quasars exhibit a broad range of molecular-gas fractions even at fixed stellar mass. {Notably, Figure~5 of \citet{Shangguan_2020} shows a weak but statistically significant correlation between molecular-gas mass and AGN luminosity (Kendall \(\tau\sim0.4\); \(>3\sigma\)), so the coupling is weak rather than absent; related galaxy-integrated diagnostics likewise indicate only a comparatively weak trend with instantaneous AGN strength \citep{Shangguan_2020_ApJ_feedback,Molina_2023}.}
These observations have led to the argument that quasar-mode feedback may appear weak or absent when assessed solely on the basis of instantaneous gas content and AGN luminosity \citep{Shangguan_2018,Shangguan_2020,Shangguan_2020_ApJ_feedback}.

{To understand these observational results,
\citet{Ward_2022} compared IllustrisTNG, EAGLE, and SIMBA with observations at
\(z\sim0\) and \(z\sim2\), and found that luminous AGN
preferentially reside in gas-rich, star-forming galaxies, with no strong negative
trend between AGN luminosity and the molecular gas fraction.
Related comparisons between cosmological simulations and observations at cosmic noon include \citet{Bertola_2024}. But due to the low resolution, the AGN physics implemented in cosmological simulation suffers from some drawbacks, including the poor determination of the black hole accretion rate (see \citet{he2025} for a detailed discussions).}

In this paper, we address this puzzle using high-resolution hydrodynamic simulations performed within the MACER framework \citep{2018ApJ...857..121Y,2025ApJ...985..178Z}. MACER is designed to simulate the evolution of individual galaxies, with particular emphasis on AGN feedback, while incorporating cosmological inflows, star formation, and stellar feedback. Three distinctive features of the MACER model, relative to other AGN-feedback models, are briefly introduced in \S\ref{sec:AGNphysics}. MACER has been applied to a range of systems and scientific questions, including the physical mechanisms responsible for maintaining massive elliptical galaxies \citep{2023MNRAS.525.4840Z}, the enhancement of star formation by AGN feedback in starburst dwarf galaxies \citep{2026ApJ..1004..128S}, and the suppression of cooling flows in galaxy clusters through the generation of strong turbulence driven by jet-wind shear \citep{he2025}. More recently, MACER has been used to systematically investigate the role of AGN feedback in the evolution of disk galaxies, with the results to be presented in a series of papers. \citet{Zou_2026} (hereafter \citetalias{Zou_2026}) provides a global overview of this series. As the second paper in the series, \citet{Zou_2026II} (hereafter \citetalias{Zou_2026II}) uses the simulation data from \citetalias{Zou_2026} to predict X-ray surface-brightness profiles and compare them with eROSITA observations. \citetalias{Zou_2026II} finds that the simulations from \citetalias{Zou_2026} reproduce the eROSITA observations remarkably well without any parameter adjustments.

As the third paper in this series, our primary goal is to test whether an ejective-feedback scenario can be compatible with gas-rich quasar hosts. Our analysis makes use of the simulation data produced in \citetalias{Zou_2026}. We compute (i) the gas-to-stellar mass ratio, \(M_{\rm gas}/M_\star\), and (ii) the cold-gas fraction, \(M_{\rm cold}/M_{\rm gas}\), and examine their relationships with instantaneous AGN activity. We then relate AGN variability to the longer-timescale evolution of the gas reservoir and the onset of quenching, thereby explaining why quasar hosts can appear gas-rich even when feedback is ultimately effective.

The paper is organized as follows. Although a full description of the simulation models is presented in \citetalias{Zou_2026}, for the convenience of the reader, we describe the salient features of the \citetalias{Zou_2026} simulation model in Section~\ref{sec:methods}. Simulation results from \citetalias{Zou_2026} that are needed to interpret the results presented here, including the star-formation and AGN light curves, are briefly summarized in the Appendix. Section~\ref{sec:results} presents the time evolution of the gas fractions and their relationships with AGN luminosity, as obtained from analysis of the simulation data. Section~\ref{sec:implication} discusses the interpretation and implications of these results in terms of the relevant timescales and the cumulative nature of ejective feedback. Section~\ref{sec:summary} summarizes the paper and discusses several caveats.

\section{The Model}
\label{sec:methods}

Our analysis in this work is based on the simulation data from the fiducial model presented in \citetalias{Zou_2026}. Below, we provide a brief summary of this model; a complete description is given in \citetalias{Zou_2026}. Appendix~\ref{app:macer} presents the star-formation and AGN-luminosity histories of the fiducial model, which provide the context for the gas-fraction analysis in Section~\ref{sec:gasfractions}.

\subsection{Simulation Setup and Initial Conditions}
\label{sec:setup}

{The simulations are performed in two-dimensional, axisymmetric spherical coordinates \((r, \theta, \phi)\) using the ZEUS-MP/2 code \citep{Hayes_2006}.
This simplification is adopted so that we can resolve the large dynamic range from the black-hole accretion scale to the circumgalactic medium, while evolving the galaxy for many gigayears with self-consistent AGN--ISM coupling; a fully three-dimensional treatment at the same resolution and duration would be prohibitively expensive.
Full technical details are given in \citetalias{Zou_2026}.}
The radial grid extends from \(100\,\mathrm{pc}\) to \(500\,\mathrm{kpc}\) and comprises 280 logarithmically spaced zones. The angular grid comprises 72 uniformly spaced zones in \(\theta\), excluding a narrow cone around the polar axis.

The galaxy model includes a central SMBH, a stellar bulge, a stellar disk, a gaseous disk, and a dark matter halo. The initial gas disk follows an exponential density profile with a scale length of \(3.5\,\mathrm{kpc}\) and a scale height of \(0.325\,\mathrm{kpc}\), and is truncated at \(R_{\mathrm{max}} = 24\,\mathrm{kpc}\) and \(|z|_{\mathrm{max}} = 2\,\mathrm{kpc}\). The initial CGM mass is \(2.0 \times 10^{10}\,M_{\odot}\). The gas is initialized with temperatures of \(10^4\,\mathrm{K}\) in the ISM and \(10^6\,\mathrm{K}\) in the CGM. The dark matter halo has a mass of \(1.6 \times 10^{12}\,M_{\odot}\) and a concentration parameter of \(c = 12\). The masses of the bulge, stellar disk, and gas disk are \(1.5 \times 10^{10}\,M_{\odot}\), \(4.7 \times 10^{10}\,M_{\odot}\), and \(0.9 \times 10^{10}\,M_{\odot}\), respectively. The initial black-hole mass is \(5 \times 10^7\,M_{\odot}\). The system is first relaxed for \(0.56\,\mathrm{Gyr}\) and subsequently evolved for \(12\,\mathrm{Gyr}\).

\subsection{Cosmological Inflows}

Cosmological inflows are modeled through hot- and cold-mode accretion, with a total inflow rate of \(100\,M_{\odot}\,\mathrm{yr}^{-1}\) divided between the cold and hot modes in a 60:40 ratio. This choice is motivated by the transitional halo mass at \(z \sim 1.5\) \citep{Dekel2009}. The hot-mode inflow is injected quasi-spherically at \(500\,\mathrm{kpc}\) with a temperature of \(10^6\,\mathrm{K}\). The cold-mode inflow is introduced through three filaments at \(100\,\mathrm{kpc}\), with a temperature of \(1.1 \times 10^4\,\mathrm{K}\) and a radial inflow velocity of \(-200\,\mathrm{km\,s}^{-1}\). In the fiducial model, the filaments are assigned a positive azimuthal velocity of \(200\,\mathrm{km\,s}^{-1}\), corresponding to high angular momentum.

\subsection{Angular Momentum Transport}

{As noted above, because our simulations are two-dimensional,} it is difficult to model angular-momentum transport processes in the galaxy self-consistently. We therefore model angular-momentum transport by introducing an anomalous stress tensor characterized by a constant kinematic-viscosity parameter, \(\nu\), which is calibrated to reproduce the inflow rates found in three-dimensional simulations of gravitationally unstable disks \citep{Hopkins2010}. The fiducial model adopts \(\nu = 0.55\), yielding an inflow rate of several tens of \(M_{\odot}\,\mathrm{yr}^{-1}\) at \(\sim 100\,\mathrm{pc}\) and an AGN duty cycle consistent with observations.

\subsection{AGN Feedback}
\label{sec:AGNphysics}

The inner boundary of the simulation domain lies within the Bondi radius \citep{2018ApJ...857..121Y}. Accretion onto the black hole within this radius is treated with sub-grid physics: we measure the mass inflow rate, $\dot{M}(r_{\mathrm{in}})$, at the inner boundary and compare it to a critical accretion rate, $\dot{M}_{\mathrm{c}}$, corresponding to a critical luminosity $L_{\mathrm{c}}=0.02L_{\mathrm{Edd}}$ \citep{McClintock2006, Yuan2014}:
\begin{equation}
\dot{M}_{\mathrm{c}} \equiv \frac{L_{\mathrm{c}}}{\epsilon_{\mathrm{EM,cold}}c^2},
\end{equation}
with $\epsilon_{\mathrm{EM,cold}}=0.1$. When $\dot{M}(r_{\mathrm{in}})>\dot{M}_{\mathrm{c}}$, the AGN operates in the cold (quasar) mode; otherwise it is in the hot (radio) mode.

Within the Bondi radius, gas falls in and circularizes at the circularization radius to form an accretion disk. The effective supply rate to the disk follows a relaxation equation with a free-fall timescale at $r_{\mathrm{in}}$, accounting for the lag between the inner boundary and the circularization radius \citep{2018ApJ...857..121Y}. The disk mass evolves according to the balance between this supply, black-hole accretion, and mass loss through winds; the black-hole accretion rate is obtained from the viscous inflow rate at the circularization radius minus the wind mass-loss rate (equations~6 and 14--17 of \citealt{2018ApJ...857..121Y}).

When $\dot{M}(r_{\mathrm{in}})>\dot{M}_{\mathrm{c}}$, the accretion flow is a geometrically thin, radiatively efficient standard disk. The AGN output is dominated by radiation and disk winds. Because only $\sim$10\% of quasars are radio-loud and the jet physics in this regime remains uncertain, jets are not included in the cold mode \citep{2018ApJ...857..121Y}. The wind mass flux and velocity as functions of bolometric luminosity are taken from the Suzaku sample of \citet{Gofford2015} (equations~9--12 of \citealt{2018ApJ...857..121Y}), with an angular distribution $\propto \cos^2\theta$. Bolometric luminosity is $L_{\mathrm{bol}}=\epsilon_{\mathrm{EM,cold}}\dot{M}_{\mathrm{BH}}c^2$. Radiation and winds are injected at the inner boundary and couple to the ISM self-consistently.

Below $\dot{M}_{\mathrm{c}}$, accretion proceeds through a hot accretion flow (ADAF/RIAF) inside a truncation radius $r_{\mathrm{tr}}$, with $\dot{M}(r_{\mathrm{in}})\approx \dot{M}(r_{\mathrm{tr}})$ \citep{Yuan2014}. Most of the inflowing gas is lost in a wind rather than accreted onto the black hole \citep{Yuan2015}. The hot-mode wind mass flux, velocity ($v_{\mathrm{w,hot}}\approx 0.2v_{\mathrm{K}}$ at $r_{\mathrm{tr}}$), and angular distribution follow \citet{2018ApJ...857..121Y}, based on GRMHD simulations. Radiative output uses the hot-mode radiative efficiency $\epsilon_{\mathrm{EM,hot}}(\dot{M}_{\mathrm{BH}})$ from \citet{Xie2012}. In the hot mode only, a relativistic jet is injected at the inner boundary with parameters from three-dimensional GRMHD simulations \citep{Yanghai_2021}: jet power $P_{\mathrm{j}}=\eta_{\mathrm{j}}\dot{M}_{\mathrm{BH}}c^2$, velocity $v_{\mathrm{j}}=0.5c$, semi-opening angle $\theta_{\mathrm{j}}=2.5^{\circ}$, and mass flux $\dot{M}_{\mathrm{j}}=0.35\dot{M}_{\mathrm{BH}}$. Jet--ISM interaction is computed self-consistently by the hydrodynamics.

\begin{figure*}[t]
  \centering
  \includegraphics[width=0.85\textwidth]{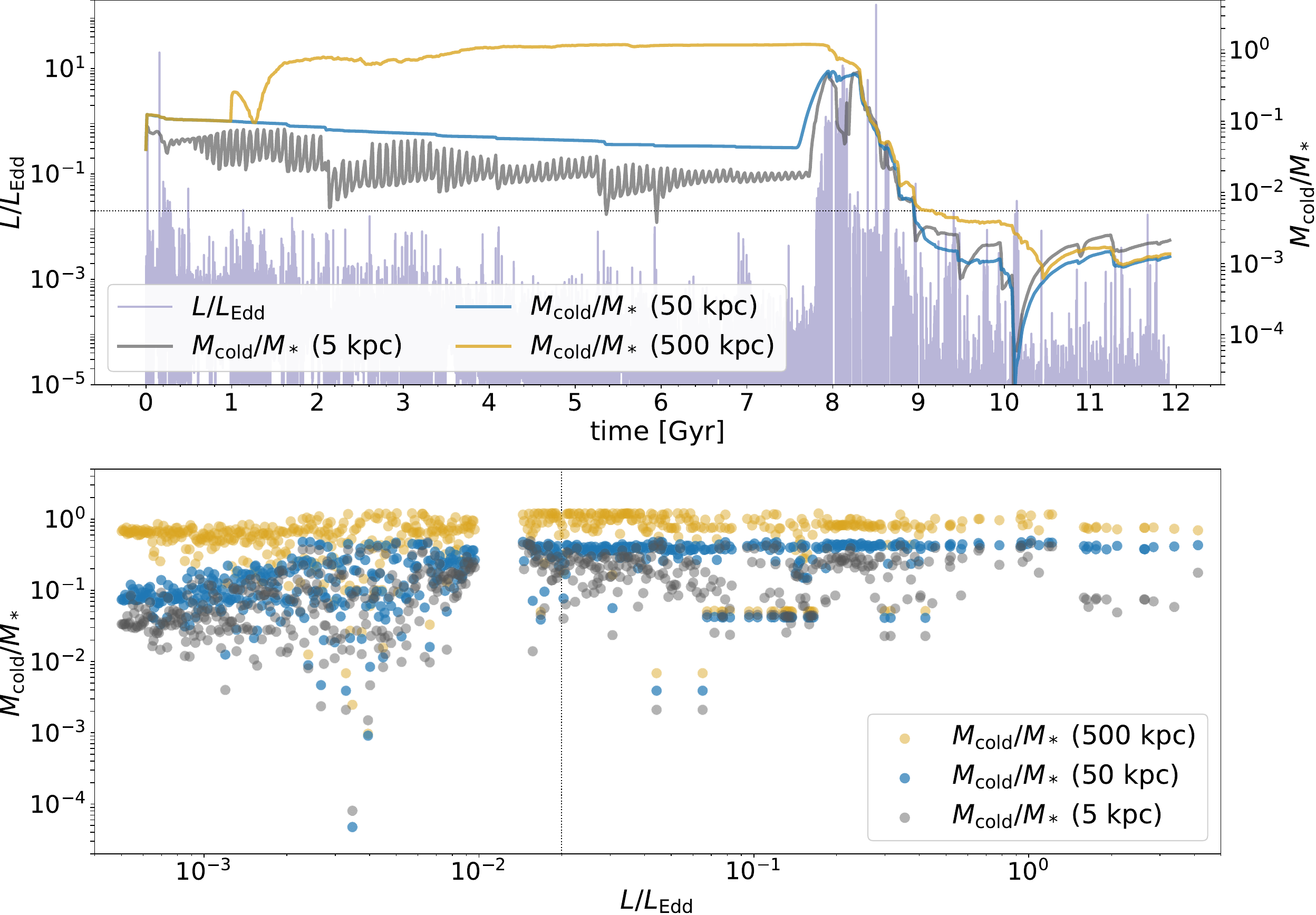}
  \caption{Evolution of the AGN Eddington ratio, $L/L_{\mathrm{Edd}}$ (purple; left axis in the top panel), and the cold-gas-to-stellar mass ratio, $M_{\mathrm{cold}}/M_{\star}$ (right axis), measured within $R=5$~kpc (gray), $R=50$~kpc (blue), and $R=500$~kpc (gold) in the fiducial model. The bottom panel shows the mean $M_{\mathrm{cold}}/M_{\star}$ in logarithmic bins of $L/L_{\mathrm{Edd}}$ for the three apertures, denoted by filled circles. The dotted horizontal line in the top panel marks $L/L_{\mathrm{Edd}}=0.02$. Except within the $R=5$~kpc aperture, $M_{\mathrm{cold}}/M_{\star}$ shows little dependence on $L/L_{\mathrm{Edd}}$, consistent with \citet{Shangguan_2018}.}
  \label{fig:BHAR-ColdGasStar}
\end{figure*}

\begin{figure}[t]
  \centering
  \includegraphics[width=\columnwidth]{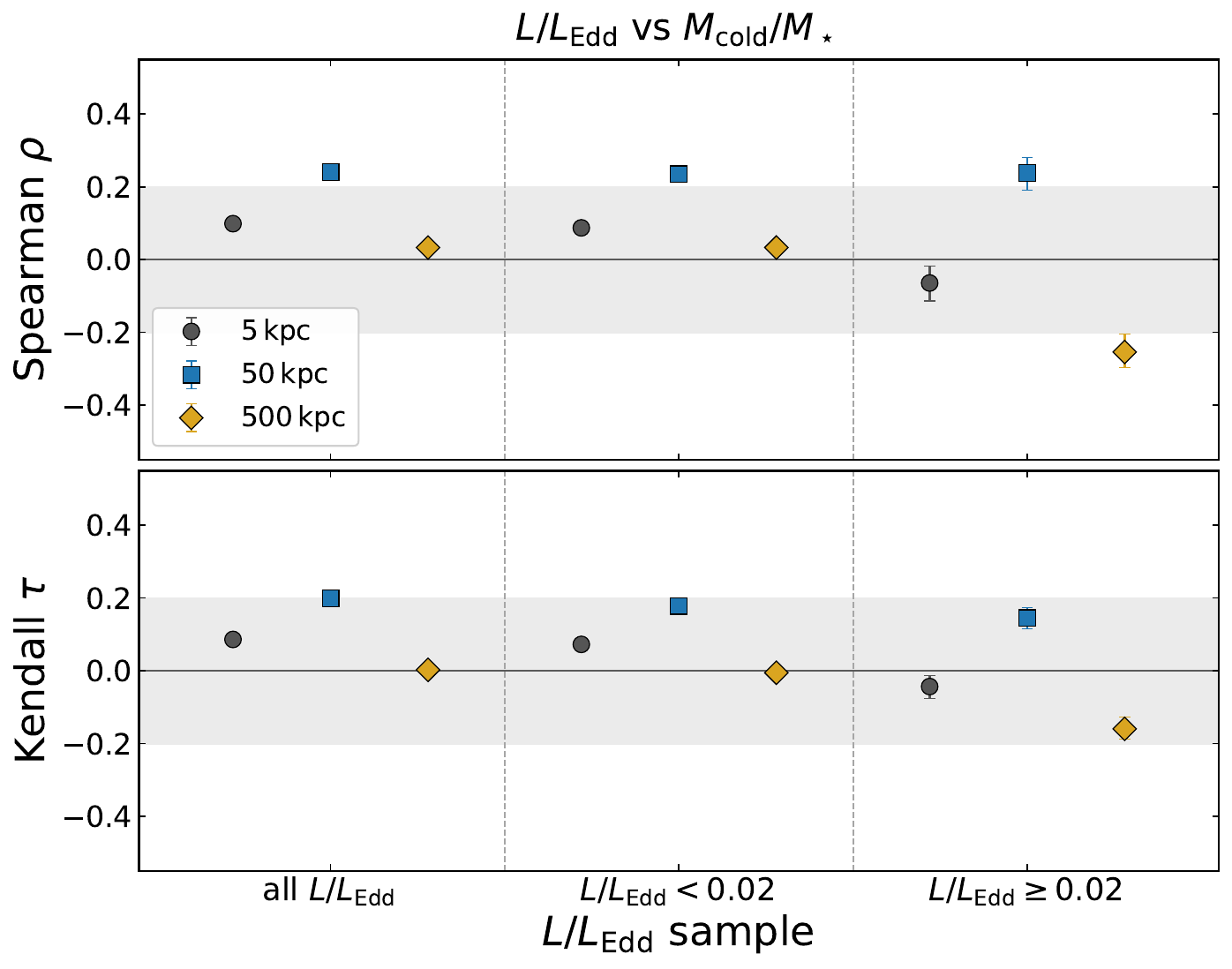}
  \caption{{Spearman \(\rho\) (top) and Kendall \(\tau\) (bottom) between \(L/L_{\rm Edd}\) and \(M_{\rm cold}/M_\star\) for three Eddington-ratio samples at \(t\geq1\,{\rm Gyr}\) in the fiducial model. Markers show the \(5\,{\rm kpc}\) (gray), \(50\,{\rm kpc}\) (blue), and \(500\,{\rm kpc}\) (gold) apertures, with the same colour coding as Figure~\ref{fig:BHAR-ColdGasStar}. Error bars denote the \(16\)th--\(84\)th percentiles from bootstrap resampling. The shaded band marks \(|{\rm corr}|<0.2\).}}
  \label{fig:corr_L_sample}
\end{figure}

\begin{figure*}[t]
  \centering
  \includegraphics[width=0.85\textwidth]{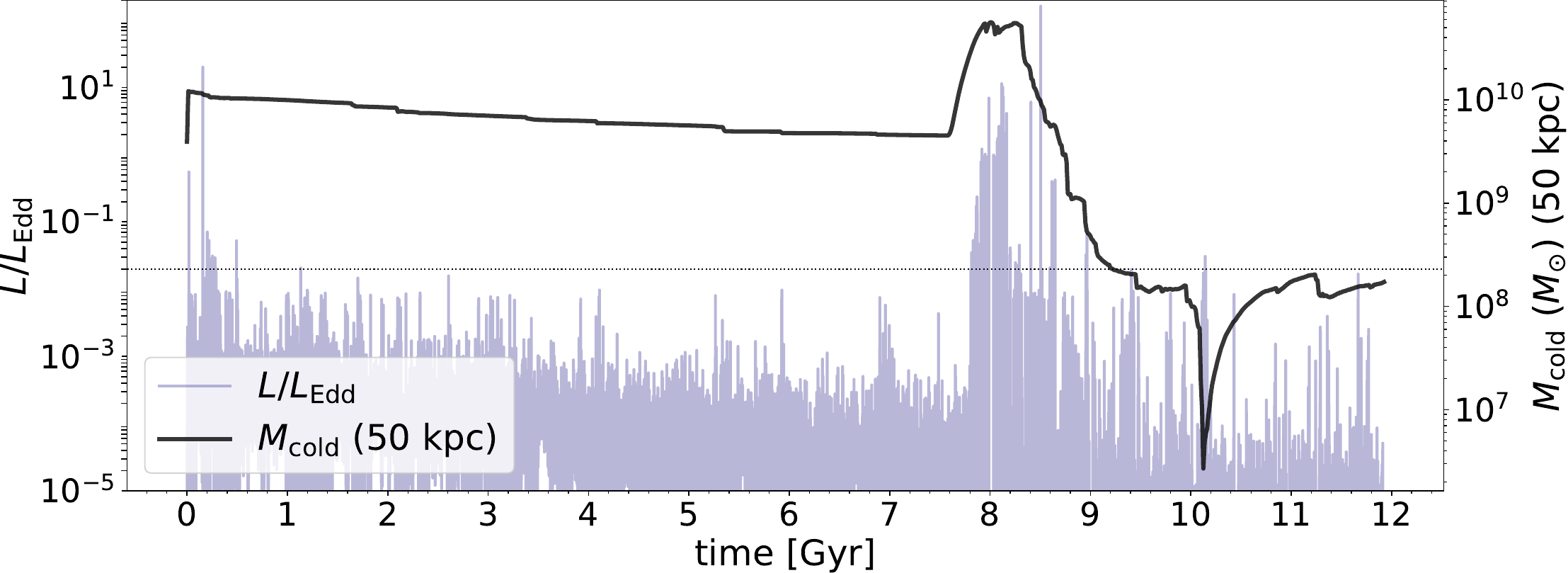}
  \caption{The purple lines show the time evolution of the AGN luminosity normalized by the Eddington luminosity, \(L/L_{\mathrm{Edd}}\) (left-hand axis). The black curve shows the time evolution of the galaxy's cold-gas mass within 50 kpc, \(M_{\mathrm{cold}}\), in the fiducial model (right-hand axis). The dotted horizontal line marks \(L/L_{\mathrm{Edd}} = 0.02\).}
  \label{fig:BHAR-ColdGasall}
\end{figure*}

\begin{figure*}[t]
  \centering
  \includegraphics[width=0.8\textwidth]{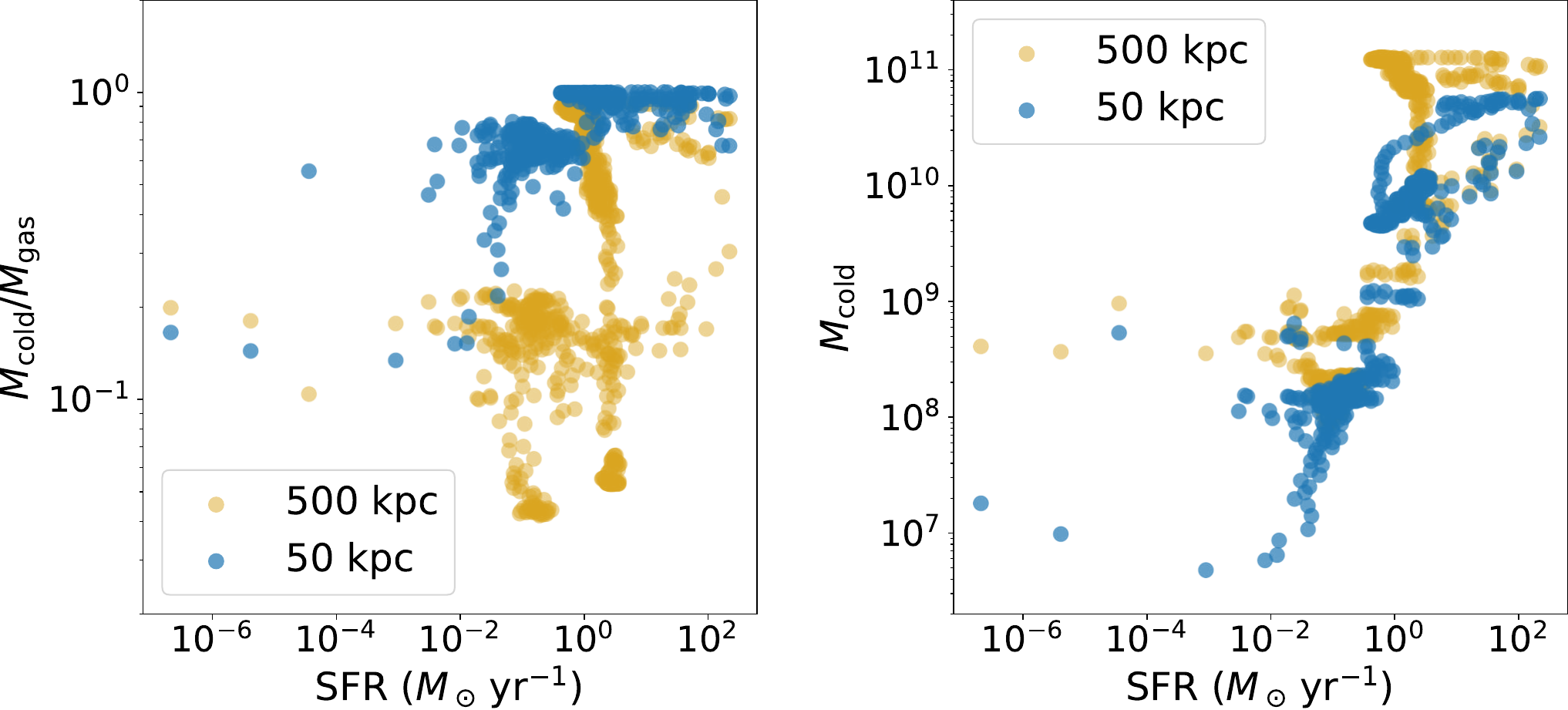}
  \caption{Correlation between the galaxy-integrated star-formation rate (SFR) and cold-gas content in the fiducial model of \citet{Zou_2026}. \emph{Left:} \(M_{\rm cold}/M_{\rm gas}\) versus SFR. \emph{Right:} \(M_{\rm cold}\) versus SFR. Filled circles denote measurements within \(R=50\)~kpc (blue) and \(R=500\)~kpc (gold), with each point representing one simulation snapshot. The right panel exhibits an approximately linear correlation, whose slope and normalization are consistent with observations.}
  \label{fig:coldgas-sfr-scatter}
\end{figure*}

\begin{figure}[t]
  \centering
  \includegraphics[width=\columnwidth]{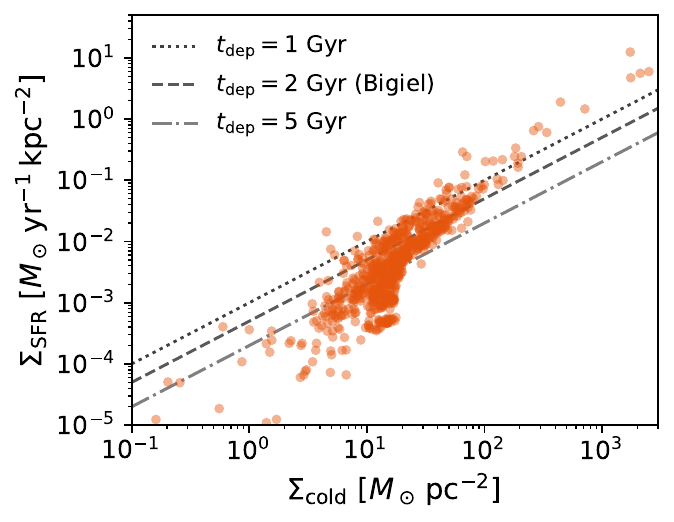}
  \caption{{Resolved Kennicutt--Schmidt-like relation in the fiducial model: cold-gas surface density \(\Sigma_{\rm cold}\) versus star-formation-rate surface density \(\Sigma_{\rm SFR}\) in midplane rings (\(|z|<1.5\,{\rm kpc}\), \(\Delta R=1\,{\rm kpc}\)) within \(R=50\,{\rm kpc}\), stacked over \(0\)--\(12\,{\rm Gyr}\). Cold gas is defined by \(T<4\times10^{4}\,{\rm K}\). Diagonal lines mark constant depletion times \(t_{\rm dep}=1\), \(2\), and \(5\,{\rm Gyr}\); the \(2\,{\rm Gyr}\) line corresponds to the characteristic molecular depletion time of \citet{2008AJ....136.2846B}.}}
  \label{fig:ks_relation}
\end{figure}

\begin{figure*}[t]
  \centering
  \includegraphics[width=0.85\textwidth]{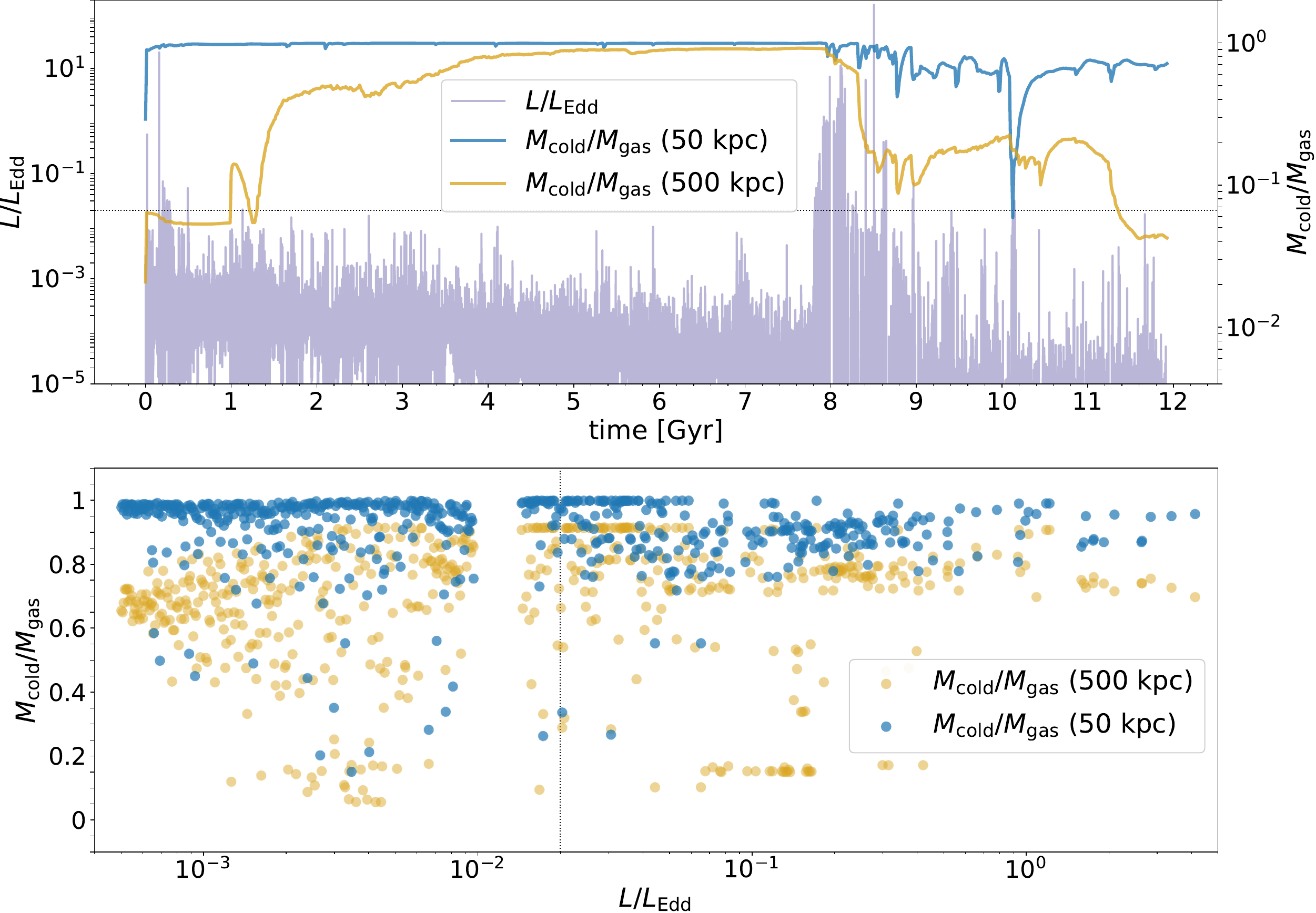}
  \caption{Evolution of the fiducial model's AGN Eddington ratio, $L/L_{\mathrm{Edd}}$ (purple; left axis in the top panel), and cold-gas mass fraction, $M_{\mathrm{cold}}/M_{\mathrm{gas}}$, measured within $R=50$~kpc (blue) and $R=500$~kpc (gold; right axis in the top panel). The bottom panel shows the mean $M_{\mathrm{cold}}/M_{\mathrm{gas}}$ in logarithmic bins of $L/L_{\mathrm{Edd}}$ for the two apertures, denoted by filled circles. The dotted horizontal line in the top panel and the dotted vertical line in the bottom panel indicate $L/L_{\mathrm{Edd}} = 0.02$.}
  \label{fig:BHAR-ColdGas}
\end{figure*}

\subsection{Star Formation and Supernova Feedback}

{Star formation is permitted independently in each grid cell with \(T < 4 \times 10^4\,\mathrm{K}\) and \(n > 1.0\,\mathrm{cm}^{-3}\). In these cells, the star-formation rate is proportional to the local cold-gas mass divided by a star-formation timescale, \(t_{\rm SF}=\max(t_{\rm cool},t_{\rm dyn})\) (equation~35 of \citealt{2018ApJ...857..121Y}), rather than by directly applying the empirical Kennicutt–Schmidt relation \citep{Kennicutt1998}. In the analysis below, we adopt the same temperature threshold to define ``cold gas.''} Supernova feedback includes both thermal and kinetic channels, with one supernova occurring per \(70\,M_{\odot}\) of stars formed. Each supernova ejects \(3\,M_{\odot}\) of gas and injects \(10^{51}\,\mathrm{erg}\) of energy, partitioned between thermal and kinetic components. The coupling radius is adjusted according to the ambient density to model the transition from the energy-conserving to the momentum-conserving phase \citep{Martizzi2015}.

\section{Results}
\label{sec:results}

\subsection{Gas-to-stellar mass ratio and gas contents}
\label{sec:gas_contents}

\citet{Shangguan_2018} investigated the cold interstellar medium in a sample of 87 low-redshift (\(z<0.5\)) PG quasars by modeling their infrared spectral energy distributions, from which they derived dust and total gas masses. They found that most quasar hosts (\(\sim90\%\)) have cold-gas fractions, defined as \(M_{\rm cold}/M_\star\), comparable to those of massive star-forming galaxies with similar stellar masses. In contrast, a minority (\(\sim10\%\)) appear genuinely gas-deficient and resemble early-type systems. The observed population spans approximately three orders of magnitude, with \(M_{\rm cold}/M_\star \sim 10^{-3}\)--\(3\). Overall, their results indicate that quasar activity does not generally lead to efficient depletion of the global gas reservoir.

We now analyze our simulation data and compare them with these observational results. The top panel of Figure~\ref{fig:BHAR-ColdGasStar} illustrates the time evolution of the AGN Eddington ratio and the cold-gas-to-stellar mass ratio, \(M_{\rm cold}/M_\star\), measured within three apertures in the fiducial model: \(R=5\,\mathrm{kpc}\), \(R=50\,\mathrm{kpc}\), and \(R=500\,\mathrm{kpc}\). The bottom panel shows the relation between these quantities. The AGN Eddington ratio exhibits rapid, large-amplitude variability, whereas \(M_{\rm cold}/M_\star\) does not track these bursts and instead evolves more gradually.

Over the full \(12\,\mathrm{Gyr}\) evolution shown in Figure~\ref{fig:BHAR-ColdGasStar}, \(M_{\rm cold}/M_\star\) within \(R=50\)~kpc---the aperture most relevant for comparison with \citet{Shangguan_2018}---spans approximately three orders of magnitude, from \(\sim10^{-3}\) to \(\sim1\), similar to the observational range reported by \citet{Shangguan_2018}. In particular, as described in Paper~I (see also Figure~\ref{fig:app_sfr_lum}), the galaxy is quenched during \(t\sim8\)--\(9\)~Gyr. Over this interval, \(M_{\rm cold}/M_\star\) declines smoothly from \(\sim1\) to \(\sim0.01\), whereas \(L/L_{\rm Edd}\) varies rapidly and does not trace this smooth evolution. This behavior is consistent with the observational results of \citet{Shangguan_2018} and \citet{Shangguan_2020_ApJ_feedback}.

Overall, our simulations support the observational finding that quasar hosts need not exhibit either elevated or suppressed \(M_{\rm cold}/M_\star\) relative to inactive galaxies, despite their substantially higher AGN activity.

{To quantify this weak coupling more directly, Figure~\ref{fig:corr_L_sample} shows Spearman and Kendall rank correlations between \(L/L_{\rm Edd}\) and \(M_{\rm cold}/M_\star\) for three Eddington-ratio samples at \(t\geq1\,{\rm Gyr}\) (the full sample; \(L/L_{\rm Edd}<0.02\); and \(L/L_{\rm Edd}\geq0.02\)), using the same three apertures as in Figure~\ref{fig:BHAR-ColdGasStar}. The correlations are generally weak: the clearest signal is a mild positive correlation at \(50\,{\rm kpc}\) (Spearman \(\rho\sim0.24\)), while \(5\,{\rm kpc}\) is weaker and \(500\,{\rm kpc}\) is essentially flat. The low-Eddington subset closely tracks the full sample, and we do not find a strong anti-correlation at any aperture. These statistics therefore support the qualitative impression from Figure~\ref{fig:BHAR-ColdGasStar} that \(M_{\rm cold}/M_\star\) depends only weakly on instantaneous AGN luminosity.}

It is noteworthy from the top panel of Figure~\ref{fig:BHAR-ColdGasStar} that, whereas \(M_{\rm cold}/M_\star\) measured within \(R=50\) and \(500\)~kpc evolves relatively smoothly, the corresponding ratio within \(R=5\)~kpc fluctuates rapidly.
{These rapid fluctuations arise because the \(5\,\mathrm{kpc}\) aperture is dominated by the nuclear and inner-disk gas, which responds promptly to intermittent AGN activity, local cooling, and recurrent outflow–inflow cycles; by contrast, the larger apertures average over a much larger reservoir and therefore evolve more smoothly (see also the discussion in \S\ref{sec:Interpretation}).}

{The same hierarchy of timescales is also relevant for transition systems that may follow luminous quasar hosts along a related evolutionary pathway.}
Whereas the PG comparison above concerns \(M_{\rm cold}/M_\star\) versus instantaneous \(L/L_{\rm Edd}\), post-starburst studies typically examine molecular-gas content as a function of evolutionary stage or star-formation indicators rather than Eddington ratio.
\citet{Yesuf_2017} compiled CO measurements for 116 post-starburst galaxies spanning a range of nuclear properties, including green-valley Seyfert candidates. They reported \(M_{\rm H_2}/M_\star\sim0.03\)--\(0.3\) for Seyfert post-starburst galaxies, with most systems lying below the molecular-gas fractions of normal star-forming galaxies. They also found a strong correlation between molecular-gas fraction and mid-infrared star-formation indicators, rather than a one-to-one correspondence with AGN class. Based on their sample and earlier CO studies of post-starburst galaxies, \citet{Yesuf_2017} argued that a single AGN episode is unlikely to remove the entire molecular-gas reservoir and that migration to the red sequence may require multiple feedback episodes over \(\gtrsim1\)~Gyr. This picture is qualitatively consistent with our finding that the galaxy-scale cold-gas reservoir can remain substantial while \(L/L_{\rm Edd}\) varies rapidly. This supports the broader conclusion that short-lived nuclear diagnostics need not trace the galaxy-integrated cold-gas reservoir on gas-removal timescales.

In addition to the gas-to-stellar mass ratio, Figure~\ref{fig:BHAR-ColdGasall} displays the evolution of the galaxy's absolute cold-gas mass alongside its AGN activity.

The purple lines trace the AGN luminosity normalized by the Eddington luminosity, \(L/L_{\mathrm{Edd}}\) (left-hand axis), while the black curve shows the cold-gas mass, \(M_{\mathrm{cold}}\), within \(50\,\mathrm{kpc}\) (right-hand axis). Observationally, the relevant ``cold-gas'' mass is typically the molecular-gas mass inferred from low-\(J\) CO-line measurements using a CO-to-\(\mathrm{H}_2\) conversion factor and, in some cases, constrained further by dust-based measurements. This quantity provides a practical estimate of the cold ISM reservoir in quasar hosts. The agreement is encouraging: in the fiducial model, \(M_{\mathrm{cold}}\) lies predominantly in the range \(10^{8}\,M_{\odot}\) to \(10^{11}\,M_{\odot}\), essentially matching the range reported by \citet{Shangguan_2019}. This result indicates that, despite large short-timescale excursions in \(L/L_{\mathrm{Edd}}\), the galaxy-scale cold-gas reservoir can remain within the broad observationally allowed range during both active and inactive phases.

\subsection{Relationships between gas content, gas fraction, and star formation rate (SFR)}
\label{sec:correlation}

As shown in \citetalias{Zou_2026}, the galaxy in the fiducial model becomes quenched after \(t\sim 8\)~Gyr, as indicated by the substantial decline in its star-formation rate (SFR; see the top panel of Figure~\ref{fig:app_sfr_lum}). Although \(M_{\rm cold}/M_\star\) shows little dependence on AGN luminosity (Figure~\ref{fig:BHAR-ColdGasStar}), one may still ask whether cold-gas content correlates with the SFR. Figure~\ref{fig:coldgas-sfr-scatter} addresses this question.

The left panel shows the relation between the cold-to-total gas mass ratio and the SFR. Only a weak correlation is found between these quantities. By contrast, the right panel shows the relation between the total cold-gas mass and the SFR in the fiducial model. This correlation is substantially stronger and approximately linear.

This stronger correlation is expected given the star-formation prescription adopted in MACER, in which the SFR is proportional to the cold-gas mass divided by a star-formation timescale (equation 35 of \citealt{2018ApJ...857..121Y}). The predicted strong linear relation between the SFR and total cold-gas mass is consistent with observations \citep{2008AJ....136.2846B,2013AJ....146...19L,2018ApJ...853..179T}. For example, \citet{2008AJ....136.2846B} presented a comprehensive analysis of the relation between the star-formation-rate surface density and gas surface density at sub-kpc resolution for a sample of 18 nearby galaxies. They found an approximately linear relation between the SFR and molecular-gas surface density, with an average depletion time of \(\sim 2\times10^9\)~yr. This result is consistent with both the slope and normalization of the relation shown in the right panel of Figure~\ref{fig:coldgas-sfr-scatter}.

{As a complementary, spatially resolved check, Figure~\ref{fig:ks_relation} shows the relation between the cold-gas surface density \(\Sigma_{\rm cold}\) and the star-formation-rate surface density \(\Sigma_{\rm SFR}\) in midplane cylindrical rings (\(|z|<1.5\,{\rm kpc}\), \(\Delta R=1\,{\rm kpc}\)) within the galaxy-scale \(50\,{\rm kpc}\) aperture, stacked over the full evolutionary track. Lines of constant depletion time \(t_{\rm dep}=\Sigma_{\rm cold}/\Sigma_{\rm SFR}\) are indicated. The simulated rings broadly occupy the same \(\sim{\rm Gyr}\) depletion-time locus as nearby star-forming galaxies \citep{2008AJ....136.2846B}. Observationally, normal disk galaxies and starbursts follow different normalizations of the gas--star formation relation, with starbursts forming stars more efficiently from a given molecular-gas reservoir \citep{Daddi_2010}. The simulation shows a similar trend across evolutionary phases: during high-SFR intervals the implied \(t_{\rm dep}\) is shorter than during more quiescent phases, consistent with a shorter \(t_{\rm SF}\) in the star-formation prescription at those times. At the highest gas surface densities, some rings deviate toward even shorter depletion times, reflecting the local dependence of \(t_{\rm SF}\) on cooling and dynamical timescales.}

\subsection{Cold-to-total gas mass ratio}
\label{sec:gasfractions}

We analyze the simulation data for the fiducial model presented in \citetalias{Zou_2026}, obtaining the evolution of the Eddington ratio, $L/L_{\rm Edd}$, and the cold-gas mass fraction, $M_{\rm cold}/M_{\rm gas}$, within two apertures:\(R=50\,\mathrm{kpc}\) (galaxy scale), and \(R=500\,\mathrm{kpc}\) (the full simulation domain). Figure~\ref{fig:BHAR-ColdGas} shows the time evolution of \(L/L_{\rm Edd}\) and \(M_{\rm cold}/M_{\rm gas}\) in the top panel, while the bottom panel presents the mean \(M_{\rm cold}/M_{\rm gas}\) in logarithmic bins of \(L/L_{\rm Edd}\).

The AGN Eddington ratio varies by more than five orders of magnitude, as the black hole alternates between luminous quasar phases and quiescent phases. In contrast, \(M_{\rm cold}/M_{\rm gas}\) remains within the relatively narrow range of \(\sim 0.1\)--\(1\) at both apertures and does not closely follow the short-timescale variability of black-hole accretion.

{Both panels of Figure~\ref{fig:BHAR-ColdGas} show that the simulated cold-gas fraction exhibits only a weak dependence on AGN activity.
As shown in \citetalias{Zou_2026} (see also Figure~\ref{fig:app_sfr_lum} in the present paper), the galaxy in the fiducial model is quenched by AGN feedback.
Therefore, a weak instantaneous correlation between the cold-gas fraction and AGN luminosity does not imply that AGN feedback is ineffective.}

\section{Implication and interpretation}
\label{sec:implication}

\subsection{Our simulations support the ejective quenching scenario}
\label{sec:ejectivequenching}

\begin{figure*}[t]
  \centering
  \includegraphics[width=1.0\textwidth]{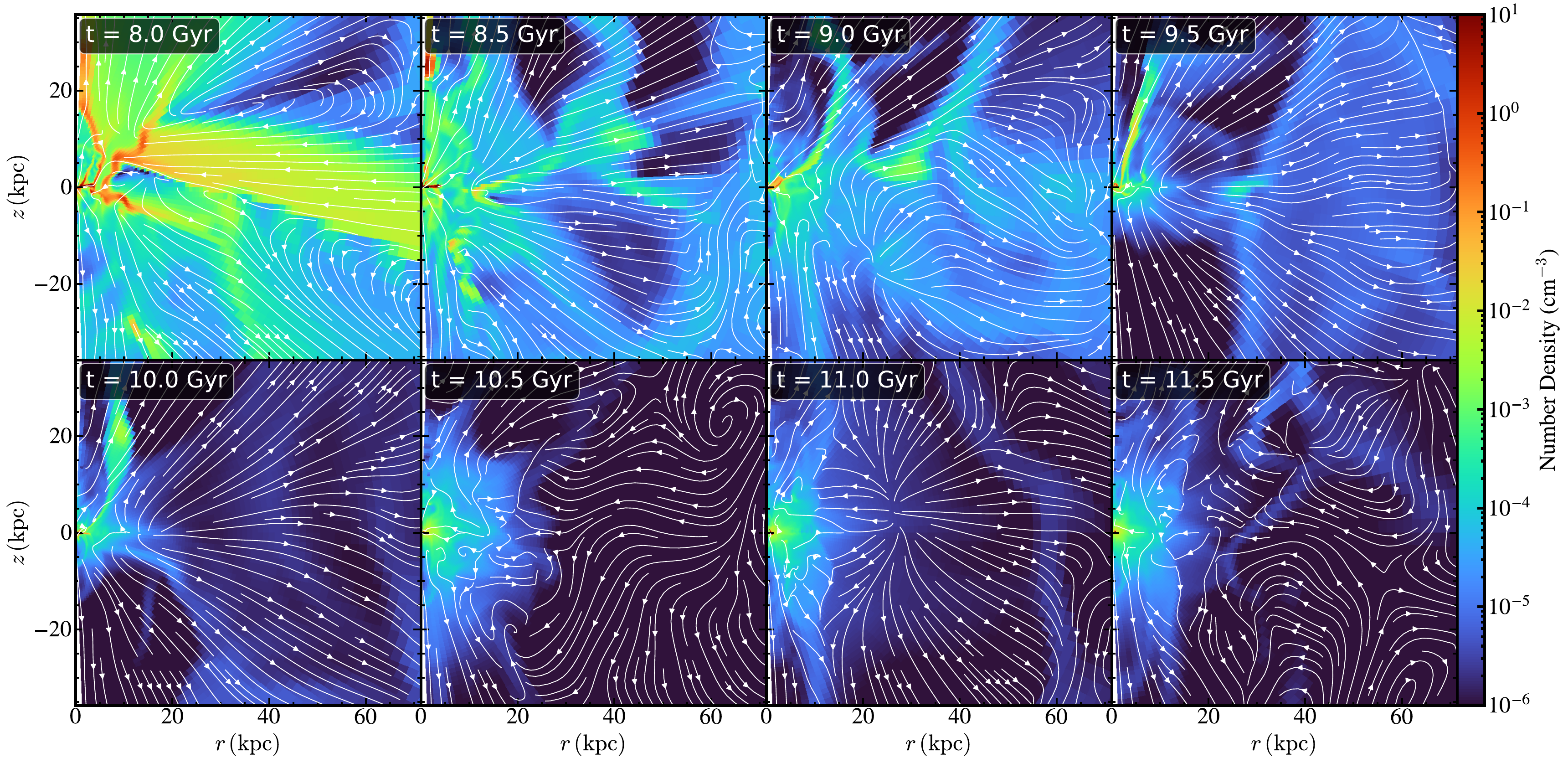}
  \caption{Evolution of gas-density distributions in the Fiducial model during the quenching phase. White arrows show the gas velocity vectors. AGN-driven outflows progressively redistribute and remove gas from the galaxy and its halo. Although some gas subsequently falls back or is replenished by cosmological inflow, the overall gas reservoir declines with time, supporting a cumulative, ejective mode of AGN feedback.}
  \label{fig:density_time}
\end{figure*}

\begin{figure}[t]
  \centering
  \includegraphics[width=\columnwidth]{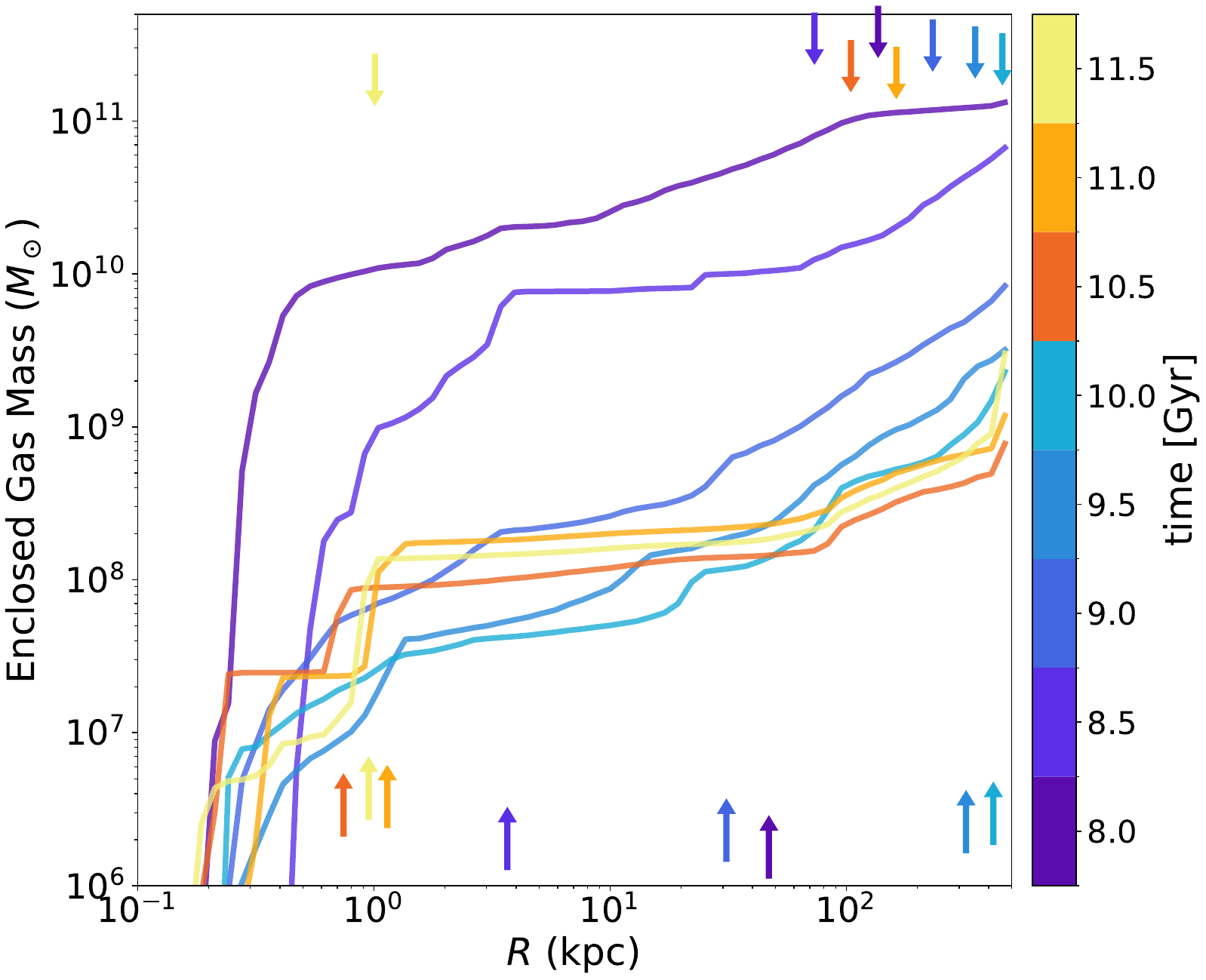}
  \caption{Cumulative gas mass enclosed within a given radius in the Fiducial model at \(t = 8.0, 8.5, 9.0, 9.5, 10.0, 10.5, 11.0,\) and \(11.5\)~Gyr.
  {Upward arrows near the bottom mark the cold-gas half-mass radius \(R_{50}\), while downward arrows near the top mark the radius enclosing 90\% of the cold-gas mass, \(R_{90}\); both use the same time color coding as the curves. Cold gas is defined by \(T<4\times10^{4}\,{\rm K}\).}}
  \label{fig:density_time_2}
\end{figure}

Our simulations support an ejective quenching mechanism.
{Here  ``ejective'' refers to AGN feedback removing a substantial amount of gas from the star-forming region of the galaxy (approximately \(R\lesssim50\,\mathrm{kpc}\) in our analysis), thereby strongly suppressing star formation to a quiescent level. This does not require that this gas be permanently unbound from the halo or escape beyond the simulation boundary at \(500\,\mathrm{kpc}\); Once displaced from the central galaxy, the gas no longer directly contributes to the star-forming reservoir, although some fraction may subsequently cool and reaccrete.}
As shown in Figure~\ref{fig:BHAR-ColdGasall}, the total cold-gas mass decreases by nearly three orders of magnitude during the quenching period, from \(t=8\) to \(9\)~Gyr. Figures~\ref{fig:density_time} and \ref{fig:density_time_2} present the two-dimensional gas-density distributions and enclosed gas-mass profiles, respectively, at eight epochs from \(t=8.0\) to \(11.5\)~Gyr, spanning the quenching phase.

The gradual decline in gas density is consistent with AGN-feedback-driven gas ejection. In particular, Figure~\ref{fig:density_time} provides direct evidence that gas is expelled from the galaxy during this period.
{Figure~\ref{fig:density_time_2} further marks the cold-gas half-mass radius \(R_{50}\) and the radius enclosing 90\% of the cold-gas mass, \(R_{90}\), at each epoch. During the strongest AGN activity and quenching interval, these sizes evolve strongly: \(R_{50}\) ranges from sub-kiloparsec to several tens of kiloparsecs at some epochs, while at others the cold reservoir becomes much more spatially extended, with \(R_{50}\) and \(R_{90}\) reaching a few hundred kiloparsecs.}

In addition to ejective quenching, another mechanism commonly discussed in the literature is preventive quenching. In this scenario, AGN activity heats the circumgalactic medium (CGM), suppresses gas cooling, and reduces the supply of cool gas available for star formation. A key expectation is that, during quenching, the cold gas is gradually consumed by star formation, whereas the hot-gas reservoir remains largely intact and its cooling is inhibited. Without replenishment of cold gas from the CGM, the galaxy should therefore exhibit a pronounced decline in its cold-gas fraction. However, as shown in the top panel of Figure~\ref{fig:BHAR-ColdGas}, the cold-gas fraction remains nearly constant over this period, particularly within \(R\lesssim50\)~kpc, where most star formation occurs.

By contrast, in our favored ejective-quenching scenario, both cold and hot gas are expelled during the quenching episode, naturally explaining the relatively modest evolution of the cold-gas fraction. The cold-gas fraction thus provides a useful, although not unique, diagnostic for distinguishing ejective from preventive quenching. Our results suggest that the galaxy is quenched primarily through ejective rather than preventive processes. A more detailed analysis of the quenching process will be presented in a subsequent paper.

\subsection{Physical interpretation: the mismatch between the timescales of AGN activity and gas removal}
\label{sec:Interpretation}

Within an ejective-quenching framework, how can we understand the simulation and observational results showing that quasar hosts have cold-gas fractions comparable to those of inactive galaxies? If a galaxy is quenched through gas ejection, one might naively expect quasar hosts to have systematically lower gas fractions.

A key point is that AGN luminosity varies on short timescales, whereas changing the galaxy-integrated gas reservoir requires a timescale that is several orders of magnitude longer. Therefore, there need not be a simple, direct correspondence between instantaneous AGN activity and the galaxy-scale gas content. The gas-ejection timescale can be estimated as the time required for AGN feedback to remove gas from the galaxy:
\[
t_{\rm ej} \sim \frac{R}{v_r}
\sim \frac{100\,\mathrm{kpc}}{100\,\mathrm{km\,s^{-1}}}
\sim 1\,\mathrm{Gyr}.
\]
{Here \(v_r\sim100\,\mathrm{km\,s^{-1}}\) is a characteristic galaxy-scale outflow/mass-transport speed for gas leaving the star-forming region. Note that this velocity need not exceed the halo escape velocity, because outflows can be continuously accelerated by the cumulative energy and momentum input from multiple AGN episodes whenever the AGN is active. The resulting \(\sim1\,\mathrm{Gyr}\) timescale therefore represents the typical depletion timescale of the galaxy-scale reservoir by many short AGN episodes, rather than a one-time expulsion of the entire ISM}\footnote{As noted above, the top panel of Figure~\ref{fig:BHAR-ColdGasStar} shows that \(M_{\rm cold}/M_\star\) measured within \(R=5\)~kpc exhibits rapid fluctuations, whereas the corresponding ratios within \(R=50\) and \(500\)~kpc evolve smoothly. The rapid variability at \(R=5\)~kpc arises because this aperture encompasses the nuclear and disk regions, where the gas responds directly to intermittent AGN activity, local cooling, and recurrent outflow–inflow cycles. By contrast, gas at \(R=50\) and \(500\)~kpc responds on much longer dynamical timescales. As AGN-driven outflows propagate outward, their short-timescale variability is smoothed by propagation delays, interactions, and mixing with the surrounding CGM, and the larger gas reservoirs at these radii.}.

By contrast, individual luminous AGN episodes are typically much shorter, lasting \(\sim10^5\)--\(10^6\,\mathrm{yr}\) in both observations and our simulations. A single bright phase therefore cannot, by itself, evacuate the galaxy-scale gas reservoir. Instead, substantial gas expulsion requires multiple cycles of AGN activity. Consequently, quasar hosts can still appear gas-rich in observations even when they are on a long-term quenching trajectory.

Within this framework, the similar gas fractions observed in quasar hosts and inactive galaxies do not contradict ejective quenching. Rather, they arise naturally from the mismatch between short-timescale AGN variability and the much slower depletion of the galaxy-scale gas reservoir.

Interestingly, the enclosed gas-mass profiles at \(t=11.0\) and \(11.5\)~Gyr (golden yellow and light yellow, respectively) lie slightly above the profile at \(t=10.5\)~Gyr over part of the galaxy, indicating that some gas is replenished after the onset of quenching.
{This suggests that AGN-driven quenching may not represent a permanent cessation of gas supply, as continued cosmological accretion and recycled gas inflows can replenish the galaxy-scale gas reservoir after AGN-driven depletion. This behavior is qualitatively consistent with Yesuf et al.~(2026, in preparation), who report widespread signatures of cool-gas inflows in transition and quiescent galaxies, suggesting that gas inflows persist even after star formation is suppressed. The possibility of post-quenching gas replenishment and galaxy revival is also consistent with recent JWST/NIRCam evidence for compact rejuvenation in a post-starburst host \citep{Zhu_2026_PhotoIFU}. In this context, quenching and inflow are not mutually exclusive, but can instead represent different stages of an ongoing baryon-cycling process.}

The detailed physical processes that determine when recurrent AGN episodes succeed in quenching the galaxy---for example, luminosity thresholds and the requirement that the cumulative AGN energy exceed the binding energy of the gas---will be examined in a forthcoming paper in this series (Zou et al. 2026, in preparation). Here we provide only a brief indication of that forthcoming analysis.

\section{Summary}
\label{sec:summary}

This is the third paper in a series of systematic studies of AGN
feedback in a disk galaxy using the MACER framework. In Paper I, we
presented an overview of the fiducial model and showed that recurrent
AGN feedback can eventually quench star formation in the galaxy. In
this paper, we have focused on an apparent observational puzzle:
low-redshift quasar hosts, in particular Palomar--Green quasars, often
contain substantial cold-gas reservoirs and exhibit cold-gas fractions
and gas-to-stellar mass ratios comparable to those of inactive
star-forming galaxies. Moreover, these galaxy-scale gas diagnostics
show only weak correlations with instantaneous AGN activity. At face
value, such observations may seem to argue against efficient AGN
feedback, especially if feedback is expected to rapidly remove the cold
gas during a luminous quasar phase.

Using the fiducial MACER simulation, which includes cosmological gas
inflow, multi-phase star formation, and two-mode AGN feedback through
radiation, winds, and jets, we have examined the relation between AGN
activity and the gas content of the host galaxy. Our main conclusions
are as follows.

\begin{enumerate}
    \item {The galaxy-scale gas content shows only a weak correlation
    with the instantaneous AGN luminosity (Figure~\ref{fig:corr_L_sample}).} In the simulation, the
    Eddington ratio varies by many orders of magnitude on short
    timescales, whereas the cold-gas fraction, $M_{\rm cold}/M_{\rm gas}$, and the gas-to-stellar mass ratio, $M_{\rm gas}/M_\star$, evolve much more gradually. As a result, active and inactive phases occupy similar regions in the gas-fraction plane. {This weak instantaneous coupling is consistent with the comparatively weak trends reported for low-redshift quasar hosts.}

\item The key physical reason is a hierarchy of timescales.
    Individual luminous AGN episodes last only $\sim 10^{5}$--$10^{6}\ {\rm yr}$, whereas the depletion of the  galaxy-scale gas reservoir by recurrent AGN-driven outflows occurs cumulatively over $\sim 1$--$2\ {\rm Gyr}$. A single AGN episode  therefore cannot remove the entire cold-gas reservoir. Instead, quenching proceeds through many cycles of accretion and feedback, so a quasar host can remain gas-rich at a given observed epoch while still being on a long-term quenching trajectory.

    \item The evolution of the gas-density distribution and enclosed
    gas-mass profiles during the quenching phase supports an ejective
    feedback picture. AGN-driven outflows progressively reduce the gas
    content of the galaxy, but the process is neither instantaneous nor
    strictly monotonic. Some gas can return or be replenished by
    cosmological inflow after the onset of quenching, indicating that
    quenching and gas inflow are not mutually exclusive. Rather, they
    are different aspects of the same baryon-cycling process.

    \item Our results favor an ejective, time-integrated interpretation
    of AGN feedback over a purely preventive scenario. In this picture,
    AGN feedback removes gas from the galaxy over many recurrent
    episodes and gradually suppresses star formation, without requiring
    the immediate disappearance of the entire cold-gas phase. This
    interpretation also provides a natural explanation for why
    post-starburst and transition systems can retain significant
    molecular gas reservoirs while evolving toward lower star-formation
    activity.
\end{enumerate}

We therefore conclude that the presence of substantial cold gas in
quasar hosts is not evidence against efficient AGN feedback. Instead,
it reflects the fact that AGN luminosity is a rapidly varying nuclear
quantity, whereas the cold-gas reservoir is a galaxy-scale component
that responds only to the cumulative effect of feedback over much
longer timescales. {Gas-rich quasar hosts are thus compatible with
ejective AGN-driven quenching, provided that feedback is understood as
a recurrent and time-integrated process rather than as a single
catastrophic blowout event.}

{The present study shows that long-term ejective feedback can coexist with substantial cold-gas reservoirs and eventually drive gas depletion in our fiducial MACER run. While this work focuses on a single, idealized evolutionary pathway rather than a full cosmological sample, the other MACER models explored in \citetalias{Zou_2026} show broadly similar behavior in the gas-fraction diagnostics considered here.}

A more quantitative comparison between the simulations and
observations will require constructing mock observational samples from
the simulation outputs. In the present work, we have compared the
intrinsic gas properties of the simulated galaxy with observed gas
fractions and gas-to-stellar mass ratios. However, observationally
inferred gas masses depend on the adopted tracers and conversion
factors, such as dust-to-gas ratios, CO-to-H$_2$ conversion factors,
and assumptions about the atomic and molecular gas components.
Moreover, observed AGN samples are selected according to luminosity,
obscuration, host-galaxy properties, and survey sensitivity. These
selection effects can influence the apparent distribution of gas
fractions and their relation, or lack thereof, with AGN luminosity.

In future work, it will therefore be important to generate synthetic observations from the MACER simulations and apply selection criteria that match those of observed quasar and AGN host samples. Such an approach would allow a direct comparison of the simulated and observed distributions of $M_{\rm cold}/M_{\rm gas}$, $M_{\rm gas}/M_\star$, star-formation rate, and AGN luminosity in the
same parameter space. It would also make it possible to test how the inferred correlations depend on observational tracers, averaging timescales, aperture effects, and sample selection. This will provide a more stringent assessment of whether the same time-scale separation identified in this work can quantitatively account for the observed gas-rich nature of low-redshift quasar hosts.
{The same framework can also help address related evolutionary questions raised by our fiducial pathway, including how common rapid AGN-driven quenching is among nearby galaxies and how such pathways relate to AGN-off/post-starburst demographics \citep{Yesuf_Bottrell_2026}. Because our simulations show that galaxies can be revived after quenching, consistent with recent JWST/NIRCam results that find compact rejuvenation in a post-starburst host after an earlier energetic phase \citep{Zhu_2026_PhotoIFU}, future mock analyses and multiwavelength comparisons will also be valuable for testing the full post-quenching evolutionary cycle.}

\appendix
\label{app:macer}
\section{Background on the Fiducial Model from Paper I}
\label{app:fiducial}

The present paper uses the fiducial run introduced and analyzed in \citetalias{Zou_2026}. The numerical setup of this model is summarized in Section~\ref{sec:methods}. Here, we provide additional background from \citetalias{Zou_2026} for readers who wish to understand the evolutionary context of the model without consulting the previous paper. We focus only on the \citetalias{Zou_2026} results that are relevant to interpreting the simulation outputs used in this work.

\citetalias{Zou_2026} presented a systematic study of AGN feedback in a disk galaxy using the MACER framework. It explored a suite of models that varied the angular momentum of cosmologically supplied gas and the efficiency of angular-momentum transport in the cold disk. The fiducial run was adopted as the reference model because it produces a realistic AGN duty cycle, recurrent luminous AGN episodes, and a transition from a star-forming galaxy to a quenched system.

\subsection{Global evolution in Paper I}
\label{app:global_evolution}

In \citetalias{Zou_2026}, the Fiducial run was shown to undergo a strong
star-forming phase followed by rapid suppression of star formation.
The star-formation rate first increases as gas accumulates and cools
in the galaxy and its surrounding circumgalactic medium. This phase is
associated with the formation of cold structures and enhanced gas
supply to the disk. The star-formation rate then reaches a peak before
declining by more than an order of magnitude as the system approaches
the quenched state.

The decline of the star-formation rate in the Fiducial run was
attributed in \citetalias{Zou_2026} to recurrent AGN feedback. Outflows powered by
the central black hole heat and displace gas from the central galaxy,
thereby reducing the fuel available for subsequent star formation.
The transition to the quenched state occurs over an extended period,
rather than through a single instantaneous event. This long-term
evolution is the reason why the Fiducial run provides a useful
baseline for studying the relation between AGN activity and host-galaxy
gas properties.

 \subsection{Star formation history relevant to this paper}
 \label{app:sfh}

\begin{figure}[t]
    \centering
    % Keep the original figure file used in the current draft.
    \includegraphics[width=0.85\textwidth]{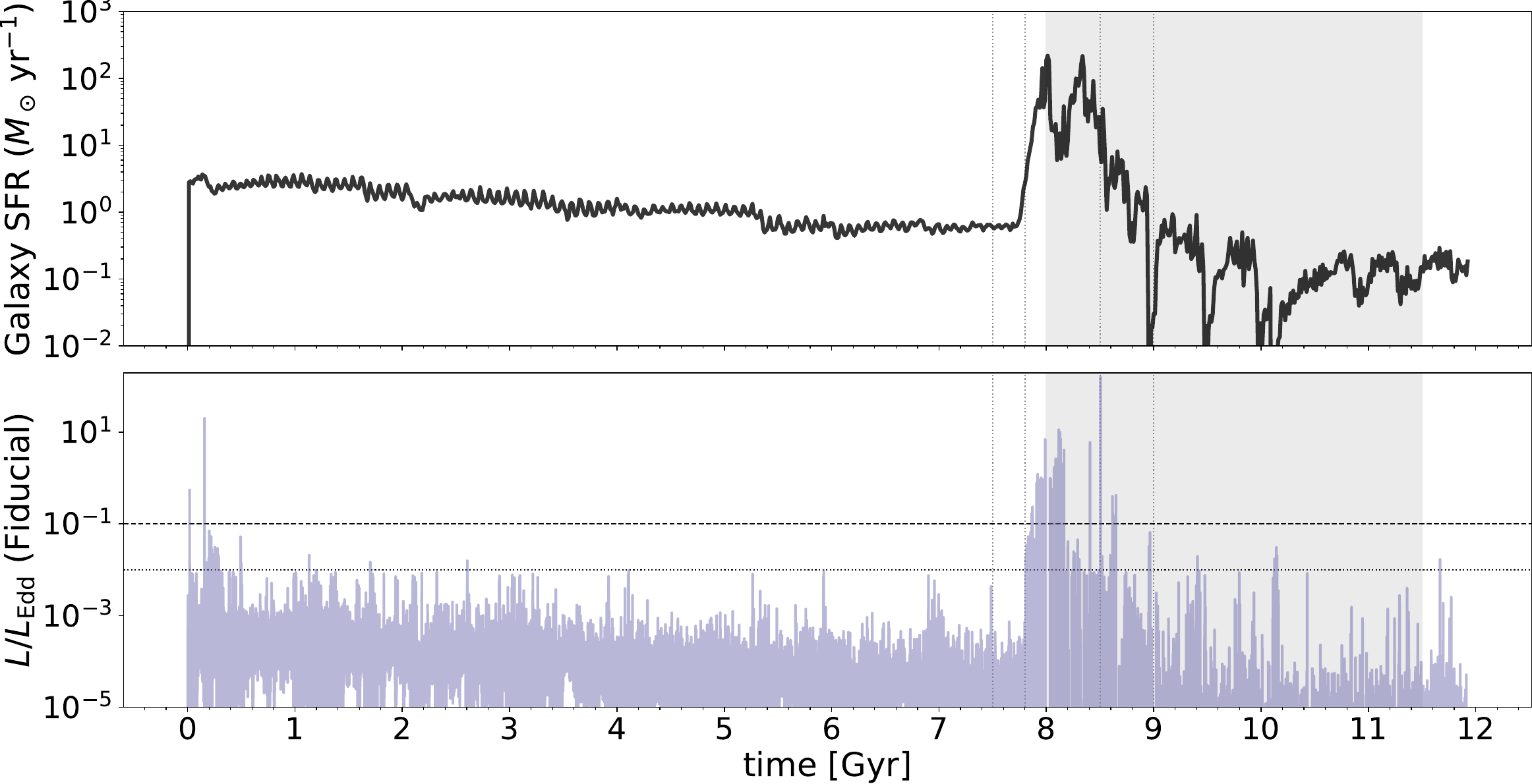}
    \caption{Star-formation and AGN luminosity histories of the Fiducial run
    from Paper I. The upper panel shows the star-formation rate, while
    the lower panel shows the AGN luminosity normalized by the
    Eddington luminosity. The shaded band marks \(t=8.0\)--11.5~Gyr (Figures~\ref{fig:density_time} and \ref{fig:density_time_2}). Dotted vertical lines indicate \(t=7.5\), 7.8, 8.5, and 9.0~Gyr. Horizontal lines in the lower panel denote \(0.01\) and \(0.1\,L_{\rm Edd}\).  This figure is included to provide background on the global evolution of the Fiducial run.    }
    \label{fig:app_sfr_lum}
\end{figure}

Figure~\ref{fig:app_sfr_lum} summarizes the two most relevant
global histories of the Fiducial run: the star-formation rate and the
AGN luminosity normalized by the Eddington luminosity. These histories
are reproduced here from \citetalias{Zou_2026} to make clear which evolutionary
phase is analyzed in the present paper.

The top panel of the figure shows that before \(t\approx7.5\)~Gyr the SFR fluctuates at modest levels; a rapid rise near \(t\approx7.5\)~Gyr marks the onset of the starburst that precedes our main analysis window. The SFR then declines sharply after \(t\approx8.5\)~Gyr, and the system is effectively quenched by \(t\approx9\)~Gyr \citepalias{Zou_2026}. The gray band highlights \(t=8.0\)--11.5~Gyr, the epoch range displayed in Figures~\ref{fig:density_time} and \ref{fig:density_time_2}.

% For the PG comparison, the important point is not the absolute SFR normalization, but the \emph{timing}: gas-fraction measurements in the main text are evaluated while the galaxy is still forming stars copiously and while nuclear activity is simultaneously strong (Section~\ref{app:agn_lum}).
 
\subsection{AGN activity and duty cycle}
\label{app:agn_duty_cycle}

\citetalias{Zou_2026} also showed that the AGN luminosity in the Fiducial run is
highly intermittent. As shown in the lower panel of
Figure~\ref{fig:app_sfr_lum}, the Eddington ratio varies by
many orders of magnitude, with short luminous episodes separated by
much longer low-luminosity intervals. Using the same definition as in
\citetalias{Zou_2026}, the AGN duty cycle of the Fiducial run is approximately
$0.49\%$, broadly consistent with observational estimates for local
galaxies hosting black holes of comparable mass.

The intermittency of the AGN activity is a direct consequence of the
self-regulated accretion and feedback cycle in the simulation. Gas
cooling and inflow can temporarily enhance the accretion rate onto the
black hole, triggering luminous AGN episodes. The resulting feedback
then modifies the gas distribution and suppresses further accretion,
after which the system may enter a low-luminosity phase until gas is
again supplied to the central region.

The late active and quenching phase of
the Fiducial run is especially useful to our present paper because it includes both active and
inactive AGN phases while the host galaxy is undergoing substantial evolution. It, therefore, provides a suitable dataset for examining how
galaxy-scale gas properties behave across different levels of nuclear activity. The detailed gas-fraction analysis based on these snapshots
is presented in the main text.  The shaded region in
Figure~\ref{fig:app_sfr_lum} indicates the approximate time
range emphasized in the present analysis.

\begin{acknowledgments}
{\it Acknowledgments} Y.Z., F.Y., and S.J. are supported by the NSF of China (grants 12192220, 12192223, 12522301, 12133008, and 12361161601), the China Manned Space Program (grants CMS-CSST-2025-A08 and CMS-CSST-2025-A10), and the National Key R\&D Program of China (No. 2023YFB3002502). Numerical calculations were run on the CFFF platform of Fudan University, the supercomputing system at the Supercomputing Center of Wuhan University, and the High Performance Computing Resource at the Core Facility for Advanced Research Computing at Shanghai Astronomical Observatory. 
\end{acknowledgments}

\section*{Data Availability}

The MACER project homepage is available at https://macer-project.github.io. The simulation data used in this paper and throughout this series are available from the corresponding authors upon reasonable request.

\bibliography{sample631}{}
\bibliographystyle{aasjournal}
\end{document}